\documentclass[10pt,final,twocolumn,journal]{IEEEtran}
\usepackage[noadjust]{cite}
\usepackage{cite}
\usepackage{graphicx}
\usepackage{caption}
\usepackage{subcaption}
\usepackage{authblk}
\usepackage{xcolor}
\usepackage{amsmath}
\usepackage{array}
\usepackage[font=small,labelfont=bf]{caption}

\begin{document}
\title{Sensitivity Challenge of Steep Transistors}
%\vspace{-3.3\baselineskip}
%\author{Authors} : Application to WSe$_2$
%Field-Effect 
\author{Hesameddin Ilatikhameneh*, Tarek Ameen*, ChinYi Chen, Gerhard Klimeck, Rajib Rahman vspace{-8ex}

%\thanks{This work was supported in part by LEAST, one of six centers of STARnet.}
%\thanks{This work was supported in part by the Center for Low Energy Systems Technology (LEAST), one of six centers of STARnet, a Semiconductor Research Corporation program sponsored by MARCO and DARPA.}
\thanks{* These authors contributed equally to this work.

This work was supported in part by the Center for Low Energy Systems Technology (LEAST), one of six centers of STARnet, a Semiconductor Research Corporation program sponsored by MARCO and DARPA.

The authors are with the Department of Electrical and Computer Engineering, Purdue University, IN, 47907 USA e-mail: hesam.ilati2@gmail.com.}
}
%the Department of Electrical and Computer Engineering, West Lafayette, , IN, 47907 USA
\maketitle
%  Center for Low Energy Systems Technology ( , a Semiconductor Research Corporation program sponsored by MARCO and DARPA
\setlength{\textfloatsep}{12pt} %{7pt plus 1.0pt minus 2.0pt}
\setlength{\belowdisplayskip}{1.6pt} 
\setlength{\belowdisplayshortskip}{1.6pt}
\setlength{\abovedisplayskip}{1.6pt} 
\setlength{\abovedisplayshortskip}{1.6pt}
%\captionsetup[table]{skip=5pt}
\setlength{\belowcaptionskip}{-12pt}
%\vspace{-1.0\baselineskip}
\begin{abstract}
Steep transistors are crucial in lowering power consumption of the integrated circuits. However, the difficulties in achieving steepness beyond the Boltzmann limit experimentally have hindered the fundamental challenges in application of these devices in integrated circuits. From a sensitivity perspective, an ideal switch should have a high sensitivity to the gate voltage and lower sensitivity to the device design parameters like oxide and body thicknesses. In this work, conventional tunnel-FET (TFET) and negative capacitance FET are shown to suffer from high sensitivity to device design parameters using full-band atomistic quantum transport simulations and analytical analysis. Although Dielectric Engineered (DE-) TFETs based on 2D materials show smaller sensitivity compared with the conventional TFETs, they have leakage issue. To mitigate this challenge, a novel DE-TFET design has been proposed and studied.   
 
\end{abstract}
\vspace{-0.5\baselineskip}
%, from transistors to photo-diodes, % at the tunnel junction %A parabolic potential profile within the depletion region at the interface.
\vspace{-0.5\baselineskip}
\section{\textbf{Introduction}}
Power consumption has been among the biggest challenges in VLSI for over fifteen years. The power dissipation, approaching the cooling limit of about 100 W/cm$^2$, has led to the saturation of frequency boost in CPUs. Decreasing the supply voltage (V$_{DD}$) to lower the power consumption in conventional MOSFETs deteriorates their switching characteristics (e.g. ON/OFF current ratio) for low V$_{DD}$ below 0.5V due to their steepness bounded to Boltzmann limit \cite{ITRS2, Mehdi}. Hence, steep transistors are critical in enabling power reduction and performance boost in CPUs and VLSIs \cite{Ionescu}. 

Tunnel FETs (TFETs) \cite{Appenzeller1, Knoch, Fiori2, Hesam1, verhulst, Fan1} and negative capacitance FETs \cite{NCFET, Zhirnov} are among the most promising candidates for steep transistors. Steepness beyond the fundamental Boltzmann limit has been experimentally observed in these devices \cite{Seabaugh, Sarkar, NC_exp}. However, there are still several challenges ahead. For example, TFET operation is based on energy filtering of hot carriers and hence, any imperfection or defect that induces states within the energy gap of semiconductor has a negative impact on its performance \cite{Avci, Esseni, SDatta}. Negative capacitance transistors usually suffer from hysteresis in transfer characteristics which increases its energy consumption, however this issue can be mitigated with proper choice of the oxide capacitance \cite{Khan}.  

These challenges in experimental realization of steep transistors obscure the fundamental issues with these steep devices. One of them is the sensitivity of device performance to its design parameters. Ideally, the steep transistor should have a high sensitivity to only the gate voltage in the subthreshold region, however, it is shown here that the conventional TFETs and NCFETs have an increased sensitivity to device parameters like the oxide and channel thicknesses and doping level. High sensitivity to variability in device parameters reduces the reliability of integrated circuits based on these devices.

\begin{figure}[!t]
        \centering
        \begin{subfigure}[b]{0.3\textwidth}
               \includegraphics[width=\textwidth]{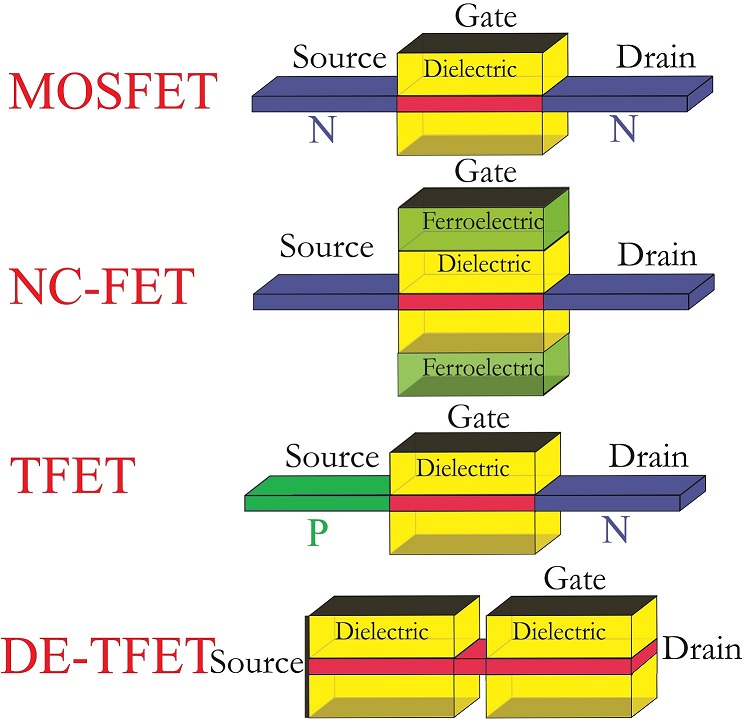}
                \label{fig:Same_EoT}
        \end{subfigure}%                                             
        %\vspace{-1.5\baselineskip}
        \caption{ \textbf{ Device structure of a MOSFET and few steep transistors discussed in this work.}}\label{fig:Fig1}
\end{figure}

To provide a quantitative measure and an intuitive insight into the sensitivity, a relative instability figure of merit $\Omega$ is defined as follows:

\begin{equation}
\label{eq:R}
I_R = \frac{I_{ON}}{I_{OFF}}
\end{equation}

\begin{equation}
\label{eq:ins22}
\Omega = \frac{ \sum\limits_{p_i \neq V_G} w_i  \frac{\partial ln(I_R)}{\partial ln(p_i)} }{  \frac{\partial ln(I_R)}{\partial ln(V_G)} }
\end{equation}

where $p_i$ is any device design parameter except the gate voltage and $w_i$ is the weighting factor that sets a weight to $p_i$ based on its importance in the sensitivity analysis. These weighting factors depend on the fabrication processes and error margins from the designed parameter space. It must be noted that if the sensitivity to device design parameters scales with the same factor as sensitivity to the gate voltage, $\Omega$ does not change. To simplify the analysis, all $w_i$s are assumed to be 1 and the device parameters are limited to source/drain doping level $N_{D}$, oxide thickness $t_{ox}$, channel thickness $t_{ch}$, dielectric constant of oxide $\epsilon_{ox}$ and channel $\epsilon_{ch}$.
\begin{equation}
\label{eq:ins2}
\Omega =  \frac{ | \frac{\partial ln(I_R)}{\partial ln(N_{D})} | + | \frac{\partial ln(I_R)}{\partial ln(t_{ox})} | + | \frac{\partial ln(I_R)}{\partial ln(t_{ch})} | + | \frac{\partial ln(I_R)}{\partial ln(\epsilon_{ox})} | + | \frac{\partial ln(I_R)}{\partial ln(\epsilon_{ch})} | }{ | \frac{\partial ln(I_R)}{\partial ln(V_G)} |}
\end{equation}

Ideally, $\Omega$ should be zero. A device with larger $\Omega$ will have more sensitivity challenges compared to those with lower $\Omega$.

In this work, the sensitivity of steep NCFET and TFET devices is studied qualitatively using analytic equations and compared to that of MOSFETs. For a quantitative analysis, full band atomistic quantum transport simulations based on non-equilibrium Green's function (NEGF) approach embedded in NEMO5 \cite{N5, N5_2} are used for MOSFETs, TFETs, and dielectric engineered (DE-) TFET \cite{Hesam3} with the structures shown schematically in Fig. 1. A novel design for DE-TFET is proposed to solve some of the challenges in the original design \cite{Hesam3}.

\section{\textbf{Analytic analysis of sensitivity}}
For MOSFETs and NCFETs, the analytic expression for current in subthreshold regime, where most of current modulation as a function of gate voltage occurs, is given approximately by:   
\begin{equation}
\label{eq:MOS}
I_{DS} = I_{OFF} ~ exp\left(  \frac{V_G}{V_{th} (1+ \frac{C_{ch}}{C_{ox}} ) } \right)
\end{equation}
where $V_{th}$ equals $K T /q$ and $I_{OFF}$ is the OFF current at $V_G$ = 0. In case of NCFETs, the effective oxide capacitance is negative which can result in an IV steeper than 60 mV/dec at room temperature if ($m = 1 + \frac{C_{ch}}{C_{ox}} < 1$):
\begin{equation}
\label{eq:SS}
SS = 60 \left(1 + \frac{C_{ch}}{C_{ox}} \right) ~ [mV/dec] = 60 m ~ [mV/dec]
\end{equation}
In case of TFETs, the approximate equation for the I-V reads as follows \cite{Analytic2, Kane1, Kane2}: 
\begin{equation}
\label{eq:MOS}
I_{DS} = I_{OFF} ~ exp\left( \Lambda \sqrt{m^*} (\sqrt{E_g + qV_G}- \sqrt{E_g}) \right)
\end{equation}
where $\Lambda$ is the total tunneling distance which is the sum of natural scaling length $\lambda$ and source depletion width $W_D$ ($\Lambda = \lambda + W_D$) \cite{Hesam_2D_Design, Analytic1}. $q$ is the unit charge and $E_g$ and $m^*$ are band gap and effective mass of the channel material. From the current equations, one can evaluate the sensitivity figure of merit $\Omega$:
\begin{equation}
\label{eq:Omega}
\Omega = \begin{cases} \frac{4}{1+ \frac{C_{ox}}{C_{ch}}}, & MOSFETs \\ |\frac{4}{m}|, & NCFET: ~ m\ll 1 \\ 1 +  \frac{3}{2} \frac{\lambda}{\Lambda}, & TFET  \end{cases} 
\end{equation}

\begin{figure}[t]
        \centering
        \begin{subfigure}[b]{0.24\textwidth}
               \includegraphics[width=\textwidth]{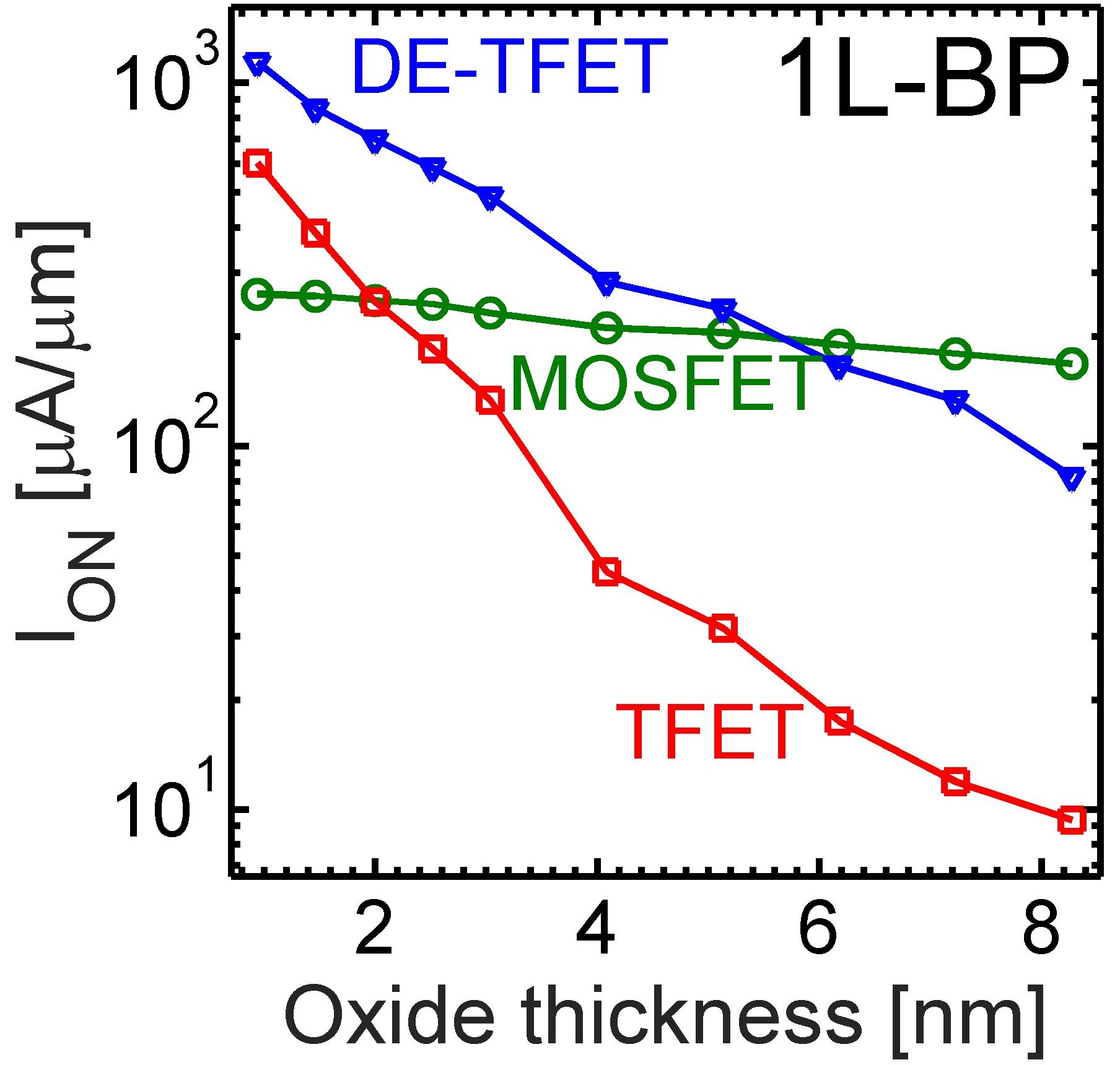}
              % \vspace{-1.5\baselineskip}
                \caption{}
                \label{fig:Same_EoT}
        \end{subfigure}%   
        \begin{subfigure}[b]{0.255\textwidth}
               \includegraphics[width=\textwidth]{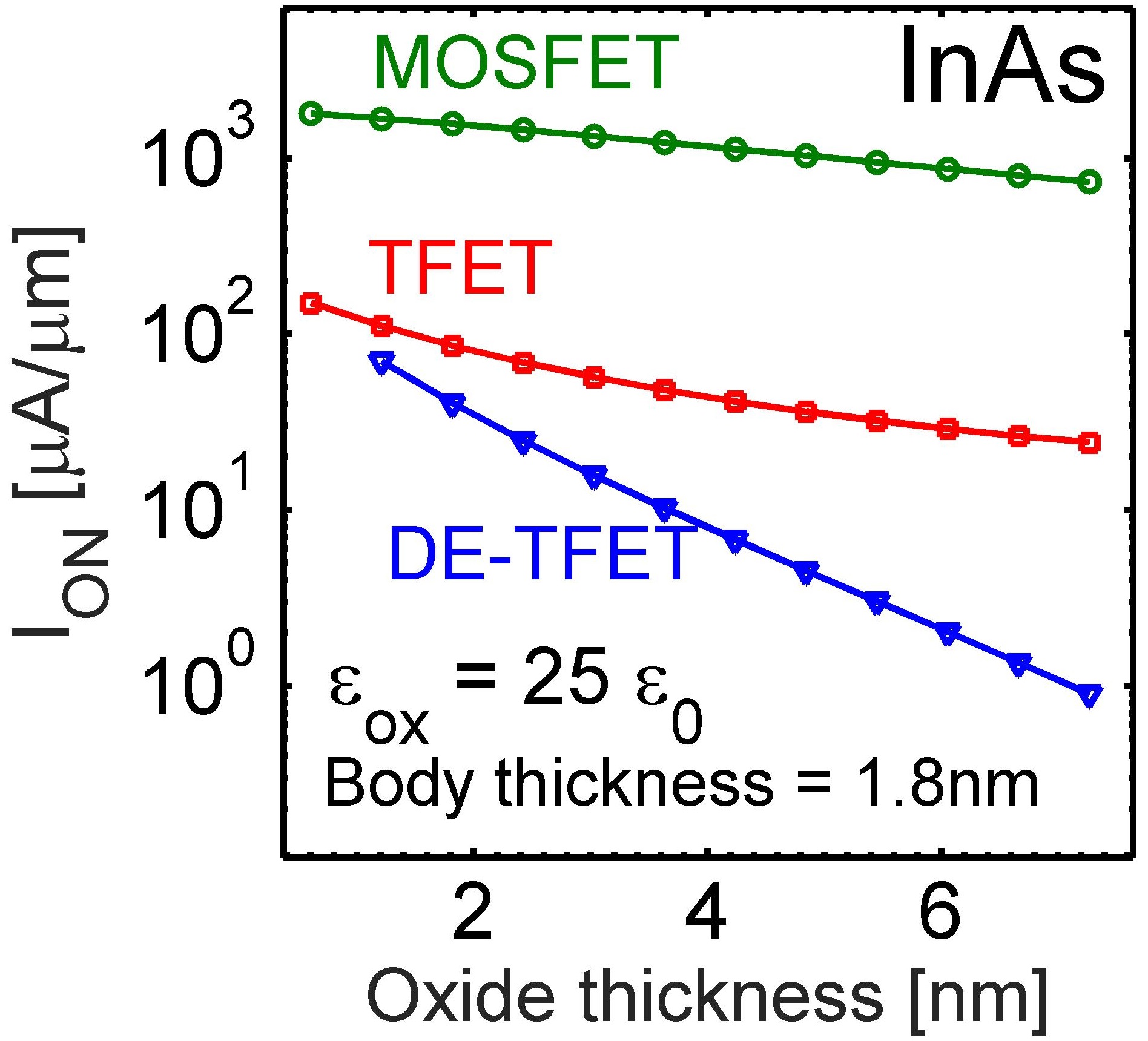}
              % \vspace{-1.5\baselineskip}
                \caption{}
                \label{fig:Same_EoT}
        \end{subfigure}%                                 
        \vspace{0.5\baselineskip}
        \caption{ \textbf{I$_{ON}$ as a function of HfO$_2$ ($\epsilon_{ox}$ = 25$\epsilon_{0}$) oxide thickness for transistors based on (a) mononlayer phosphorene and (b) InAs channel materials.} }\label{fig:Fig1}
\end{figure}

For MOSFETs, $\Omega$ reduces to 0 if $C_{ox} \gg C_{ch}$. This high stability of a MOSFET arises from the fact that for a large $C_{ox}$, the surface potential has a one to one dependency on the gate voltage, that stays intact with small variations in device design parameters. However, in case of conventional NCFETs, $\Omega$ cannot decrease since it is inversely proportional to $m$ (see equation (\ref{eq:SS})). NCFETs require a smaller $m$ to decrease SS which leads to high value of $\Omega$. Intuitively, NCFETs increase their steepness by making $\frac{C_{ox}}{C_{ch}}$ close to -1. Hence, $m$ gets close to 0 and any small variation in channel or oxide capacitances can lead to a significant change in steepness and current levels. Conventional TFETs also suffer from high $\Omega$ values and it is impossible to reduce $\Omega$ to 0 in these devices. This higher sensitivity of TFETs arises from the exponential dependence of tunneling current on tunneling distance, which depends on oxide and channel thickness and source doping level \cite{Hesam_2D_Design}.  
  
In summary, from an analytic perspective, steep transistors of conventional designs suffer from a high sensitivity to device parameters. In the next section, numerical quantum transport simulations have been used to study the impact of design parameters on the performance of steep transistors, compared to the conventional MOSFETs.
      
\section{\textbf{Atomistic quantum transport simulations}}
Two major candidates for channel material in the next generation devices are III-V \cite{Takagi_35} and 2D Van der Waals materials \cite{MoS2_Nature, Changxi, Tao}. InAs and phosphorene have been chosen as channel materials in this work due to their promising characteristics; InAs has a high mobility and geometric confinement increases the bandgap of InAs to a desirable range (above 0.7eV) \cite{Sub10}. Phosphorene has anisotropic effective mass and a necessary bandgap for TFET applications \cite{Tarek1, Phos1, Hesam4, MoS2_photo2, TETFET}. 

\begin{figure}[!t]
        \centering
        \begin{subfigure}[b]{0.24\textwidth}
               \includegraphics[width=\textwidth]{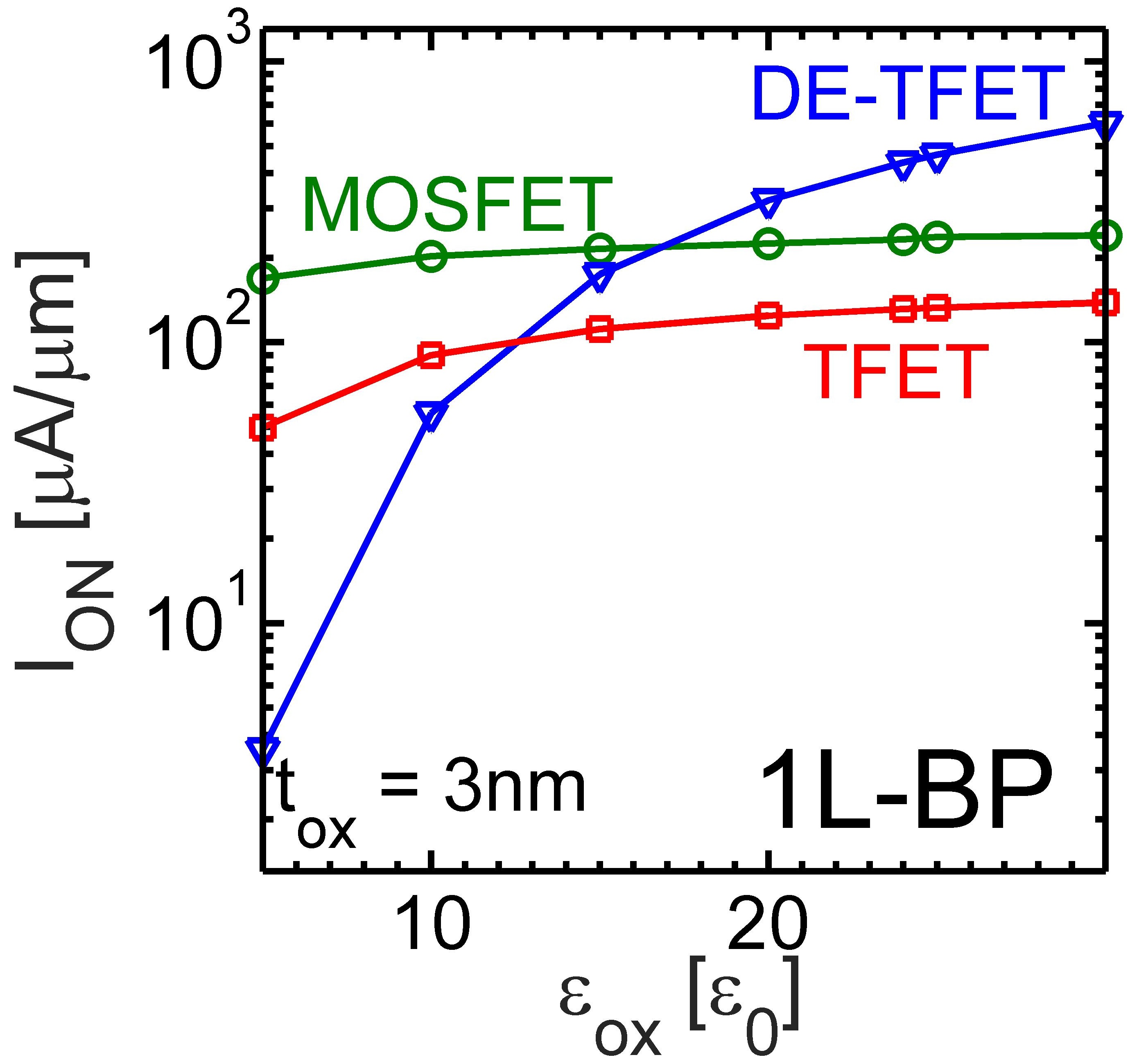}
              % \vspace{-1.5\baselineskip}
                \caption{}
                \label{fig:Same_EoT}
        \end{subfigure}%   
        \begin{subfigure}[b]{0.24\textwidth}
               \includegraphics[width=\textwidth]{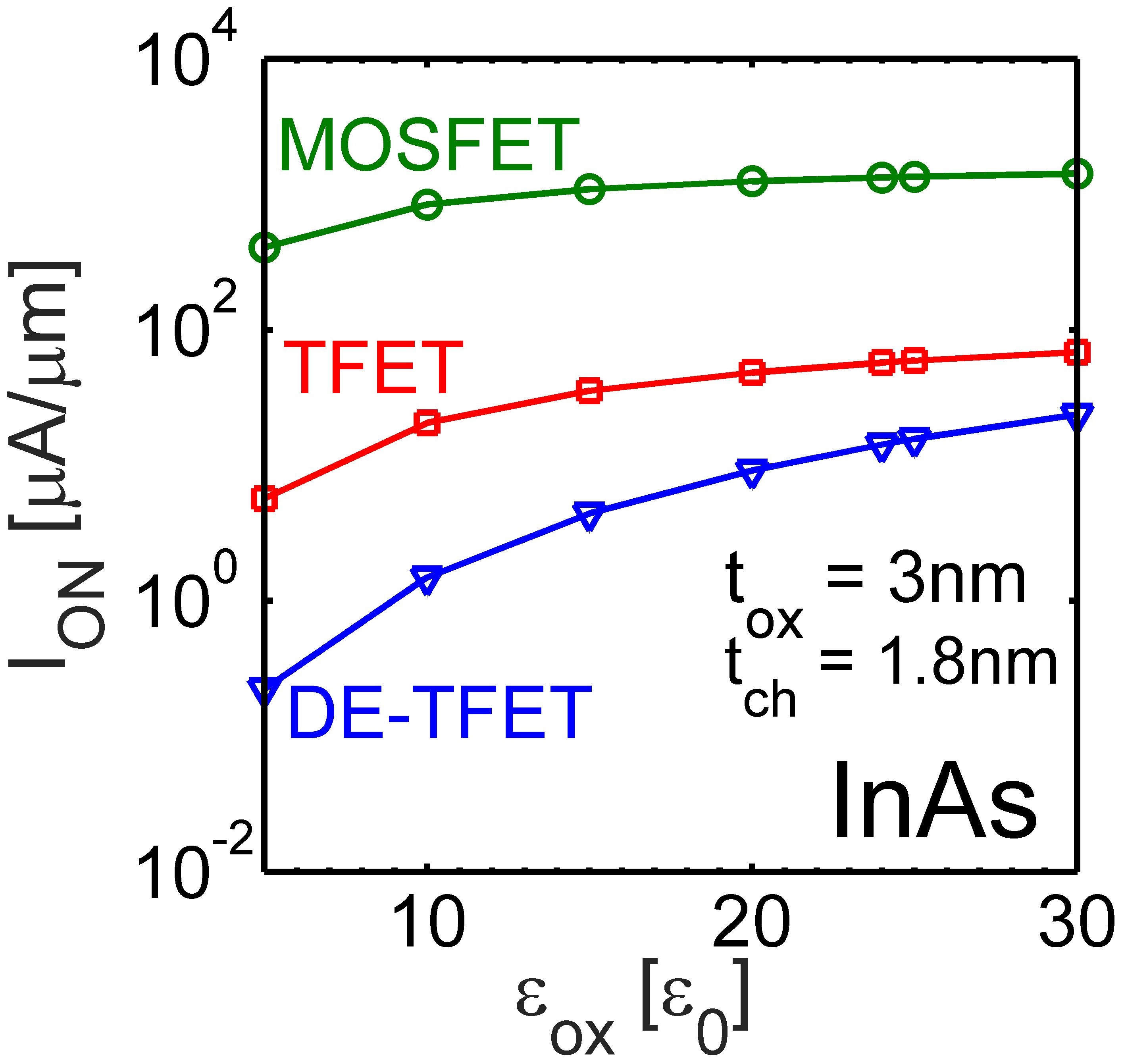}
              % \vspace{-1.5\baselineskip}
                \caption{}
                \label{fig:Same_EoT}
        \end{subfigure}%                                       
        \vspace{0.7\baselineskip}
        \caption{ \textbf{I$_{ON}$ as a function of oxide dielectric constant ($\epsilon_{ox}$) in monolayer phosphorene and InAs transistors.} }\label{fig:Fig1}
\end{figure}

\begin{table}[!b]\center
\vspace{1.0\baselineskip}
\caption{\label{tab:tran_prop} { \textbf{Default design parameters of transistors: supply voltage, channel thickness (t$_{ch}$), oxide thickness (t$_{ox}$) and dielectric constant ($\epsilon_{ox}$) and spacer's dielectric constant ($\epsilon_{S}$) and length for DE-TFET ($S$). }}}
    \begin{tabular}{| l | l | l | l | l | l | l |}
    \hline
	Material 	& $V_{DD}$ 	& $t_{ch}$	& $t_{ox}$ 	& $\epsilon_{ox}$ 	& $\epsilon_{S}$  & $S$\\ \hline
	Phosphorene & 0.5V	 	& 0.5nm 	& 3.0nm 	& 25$\epsilon_{0}$  & 1$\epsilon_{0}$ & 2nm \\
	InAs        & 0.5V 		& 1.8nm 	& 3.0nm 	& 25$\epsilon_{0}$  & 1$\epsilon_{0}$ & 2nm \\
	\hline
    \end{tabular}
\end{table}
        
Full-band atomistic quantum transport simulations based on self-consistent solution of 3D Poisson equation and non-equilibrium Green's function (NEGF) method \cite{Luisier_2006_ref} is used in numerical simulations of the devices. The Hamiltonian is represented by $sp^3d^5s^*$ tight-binding (TB) model \cite{SlaterKoster}, considering 1st nearest neighbor for InAs \cite{InAs_param} and 2nd nearest neighbor for phosphorene \cite{Tarek1}. The TB parameters are optimized based on ab-initio bandstructures obtained from density functional theory using the hybrid functionals for phosphorene (HSE06) \cite{Hesam4, Tarek1}.

A supply voltage of 0.5V has been used in all the atomistic simulations of transistors in this work. Ultra-scaled MOSFETs usually operate at a V$_{DD}$ beyond 0.5V whereas TFETs are beneficial at V$_{DD}$ less than 0.5V. Hence, we choose a value at the boundary of the 2 regions where both MOSFETs and TFETs are operational. The default transistor parameters are listed in table I.

Fig. 2a shows the dependence of I$_{ON}$ for different 2D transistors based on monolayer phosphorene as a function of HfO$_2$ oxide thickness. Following the standard practice in transistor benchmarking, I-Vs are shifted horizontally such that I$_{OFF}$ is fixed at 1$nA/\mu m$ at zero gate voltage. In case of 2D materials, DE-TFET outperforms MOSFET and TFET at smaller HfO$_2$ thicknesses. The sensitivity of DE-TFET to $t_{ox}$ variations is lower than a TFET but higher than a MOSFET. As predicted from the analytic theory, MOSFET is the most stable of the explored devices and is least sensitive to $t_{ox}$ variations.   

Fig. 2b shows I$_{ON}$ as a function of HfO$_2$ thickness for III-V transistors with an InAs channel of thickness 1.8 nm. The increase in channel thickness from 0.5nm in phosphorene to 1.8nm in InAs was enough to reduce the performance of DE-TFET below a conventional TFET. The tunneling distance in electrically doped devices \cite{Hesam2, Hesam3, Hesam_2D_Design} depends on the channel thickness. Moreover, III-V DE-TFETs and conventional TFETs not only provide less ON-current, smaller by an order of magnitude, but also suffer from an increased sensitivity to $t_{ox}$ variations. 

\begin{figure}[!t]
        \centering
        \begin{subfigure}[b]{0.24\textwidth}
               \includegraphics[width=\textwidth]{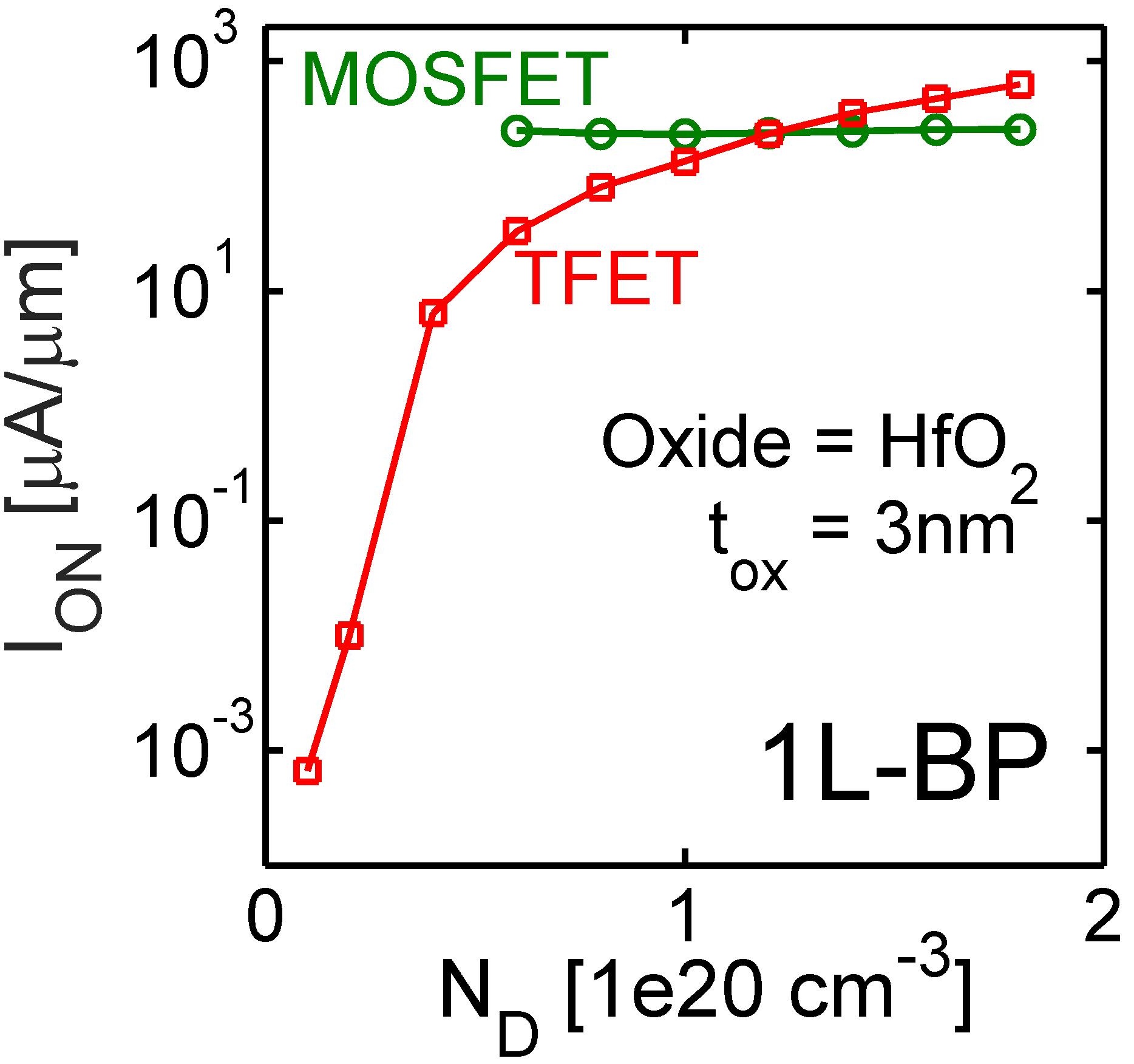} %Fig1_3
               \vspace{-1.5\baselineskip}
               \caption{}
        \end{subfigure}       
        \begin{subfigure}[b]{0.24\textwidth}
               \includegraphics[width=\textwidth]{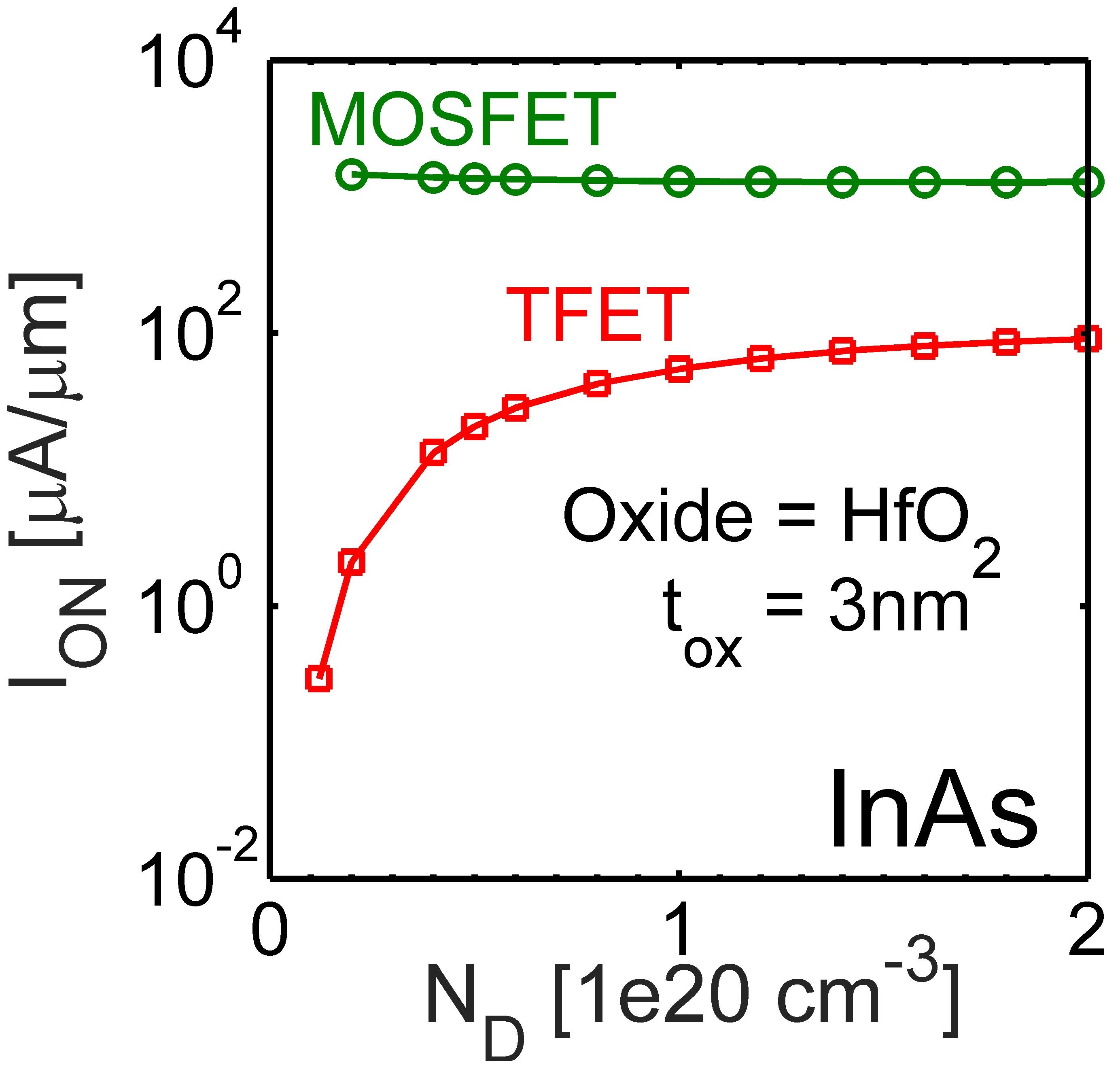} %Fig1_3
               \vspace{-1.5\baselineskip}
               \caption{}
        \end{subfigure}%                                               
        \vspace{0.7\baselineskip}
        \caption{ \textbf{I$_{ON}$ as a function of source doping level (N$_D$) in MOSFETs and TFETs based on a) monolayer phosphorene and b) InAs.} }\label{fig:Fig3}
\end{figure}

Fig. 3a shows the I$_{ON}$ of phosphorene based transistors with varying oxide dielectric constant. Although the dielectric constant of an oxide depends on the dielectric material, the environment can affect the electrostatics \cite{HfO2_1}. Since the electric field amplification at the tunnel junction in a DE-TFET depends on the ratio of high-k and low-k dielectrics, its performance is more sensitive to $\epsilon_{ox}$. This sensitivity reduces for higher $\epsilon_{ox}$ values which underlines the importance of high-k dielectrics for DE-TFETs. Nevertheless, the performance of a DE-TFET is higher than both the conventional TFET and MOSFET for a large range of $\epsilon_{ox}$. The impact of $\epsilon_{ox}$ on I$_{ON}$ for InAs transistors is shown in Fig. 3b. Again, III-V MOSFET outperforms both TFET and DE-TFET in terms of both performance and sensitivity.

Fig. 4a shows I$_{ON}$ of MOSFET and TFET as a function of source doping level in phosphorene based transistors. Since DE-TFET uses electrical doping instead of chemical doping, it is immune to dopant fluctuations. In general, MOSFET is much less sensitive to the doping level variations. For high source doping levels, TFET has smaller sensitivity to doping level variations, as compared to the low doping levels. Fig. 4b shows I$_{ON}$-N$_D$ plot for InAs based transistors which appears similar in behavior to the 2D transistors.

Fig. 5 illustrates the sensitivity of transistors to their channel thickness ($t_{ch}$) in multilayer phosphorene MOSFET, TFET, and DE-TFETs. Increasing the channel thickness reduces the bandgap and effective mass of InAs and multilayer Phosphorene channels. Due to smaller bandgap I$_{OFF}$ increases beyond 1$nA/\mu m$ for some channel thicknesses. Hence, I$_{ON}$/I$_{OFF}$ is shown in the Figs. 5a and 5b. MOSFETs have the least sensitivity to $t_{ch}$ and DE-TFETs are moderately less sensitive than TFETs. Phosphorene tunneling devices have significantly larger I$_{ON}$/I$_{OFF}$ compared to the InAs ones due to the smaller $E_{g}$. However, for larger $t_{ch}$ phosphorene’s $E_{g}$ becomes too small to suppress the OFF current for all devices. 

In summary, the quantum transport simulations show that 2D DE-TFET can outperform TFETs and MOSFETs in terms of performance. Although MOSFETs are least sensitive to design parameter variations, DE-TFET can outperform TFETs with its immunity to doping level variations, a high-k dielectric and a thin channel.    

\begin{figure}[!t]
        \centering
        \begin{subfigure}[b]{0.24\textwidth}
               \includegraphics[width=\textwidth]{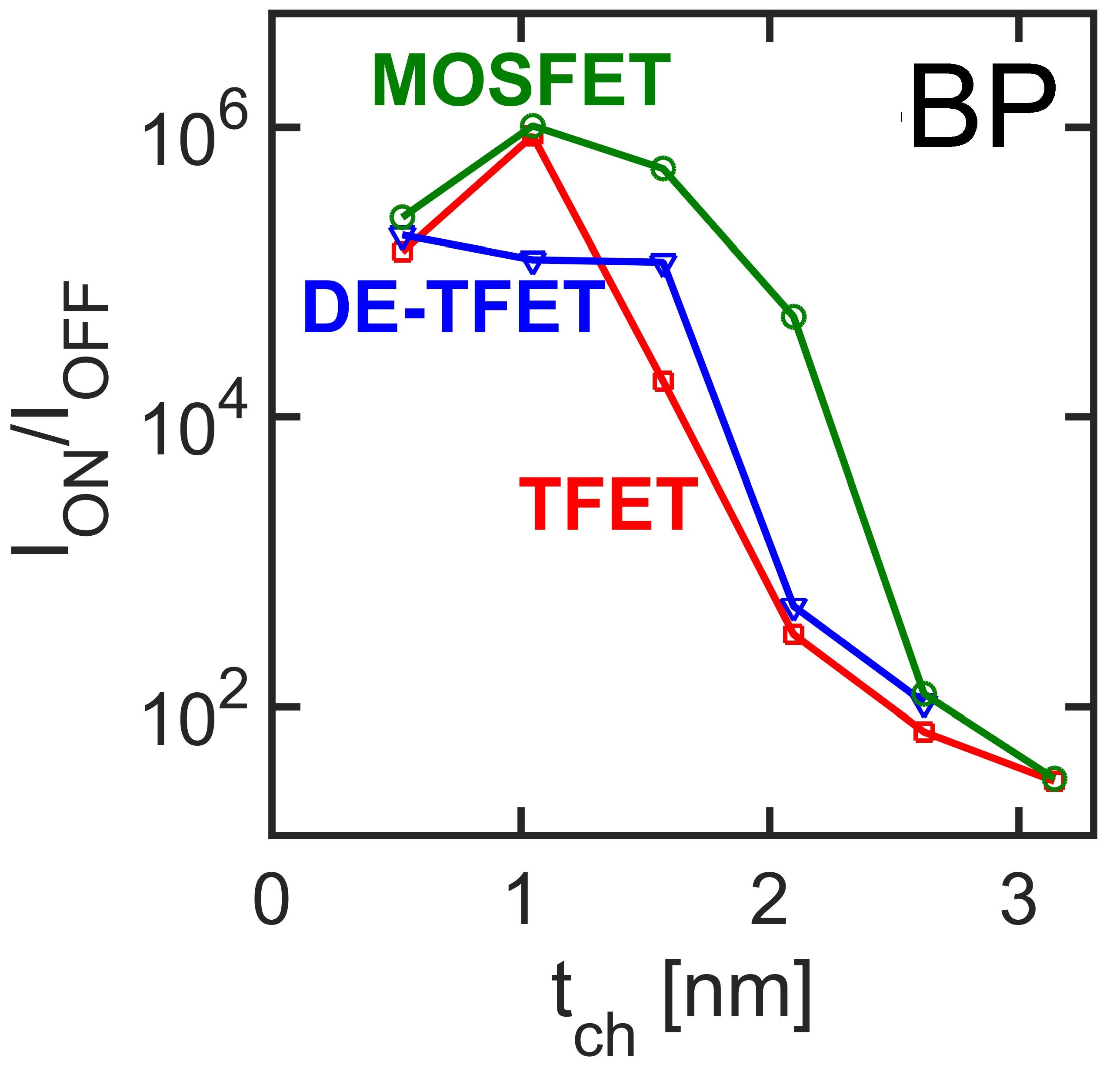}
               \vspace{-1.5\baselineskip}
                \caption{}
                \label{fig:Same_EoT}
        \end{subfigure}%                       
        \begin{subfigure}[b]{0.24\textwidth}
               \includegraphics[width=\textwidth]{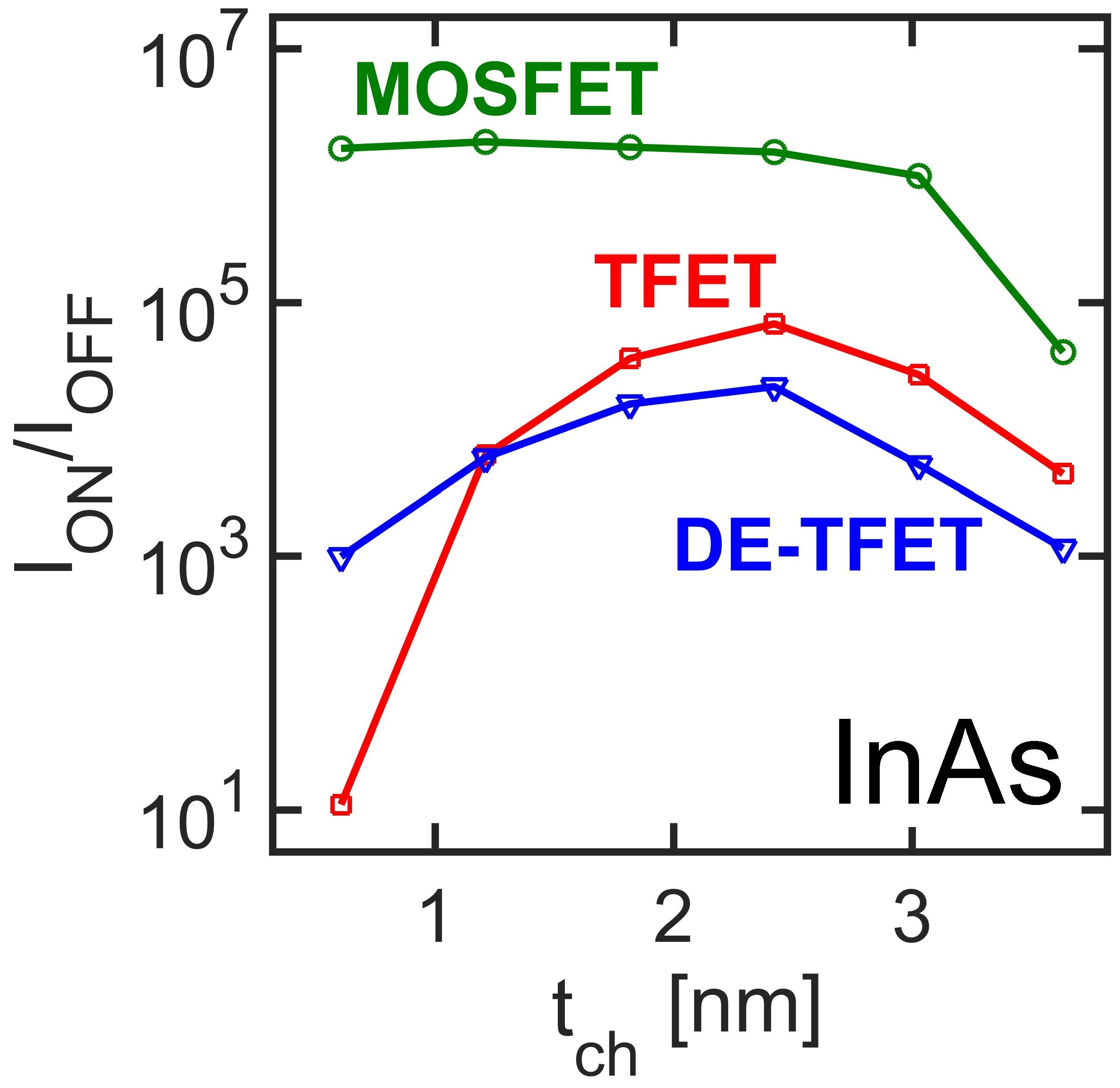}
               \vspace{-1.5\baselineskip}
               \caption{}
                \label{fig:Same_EoT}
        \end{subfigure}%          WDE_TFET                               
        \vspace{0.5\baselineskip}
        \caption{ \textbf{I$_{ON}$/I$_{OFF}$ as a function of channel thicjness ($t_{ch}$) in multilayer phosphorene and InAs transistors.} }\label{fig:Fig1}
\end{figure}

\section{\textbf{Workfunction engineering}}
Although DE-TFET is shown to outperform the MOSFET and the conventional TFET in 2D channel materials, for suitable ranges of device parameters, the leakage current between the contacts (Fig. 6a) can deteriorate its performance. The optimum spacing between contacts is a few nanometers. An attempt to decrease leakage by increasing this spacing reduces the on-current of DE-TFETs. One way to lower the current between the contacts is to reduce the difference in chemical potential between them. According to the Landauer's formula \cite{Landauer}, it is known that the leakage current between the gates ($I_{Leakage}$) 
\begin{equation} 
I_{Leakage} \propto T (\mu_1 - \mu_2) 
\label{eq:opt1}
\end{equation}
where T is the tunneling transmission and $\mu_1$ and $\mu_2$ are chemical potentials of the two gate metal contacts. Shorting the two gate metal contacts with different work functions will force $\mu_1$ and $\mu_2$ to be equal thus ensuring a zero leakage current between the gates $I_{Leakage}$. Hence, using two connected metal gates with different workfunctions in a DE-TFET, the original design is improved to work function engineered DE-TFET (WDE-TFET), shown schematically in Fig. 6b. Forcing $\mu_1=\mu_2$ also causes a potential drop between the regions under the two gates and a tunneling window is set by this difference in work functions of the two gate metals. However, in such a design with shorted gates, it appears that this tunneling window is material dependent and cannot be controlled with a gate bias (i.e. tunneling window $\approx \Phi_{M1} - \Phi_{M2} - E_g$), thus rendering the device incapable of turning OFF. But it must be noted that although the two gates are connected, the surface potential under the gate metal closer to the source does not follow the gate voltage due to large charge concentration in this region that screens out the gate voltage. Therefore, the response of the surface potential to the gate voltage is different in the regions under the two gates leading to a tunneling window that can still in effect be controlled by the gate bias. 

\begin{figure}[!t]
        \centering
        \begin{subfigure}[b]{0.27\textwidth}
               \includegraphics[width=\textwidth]{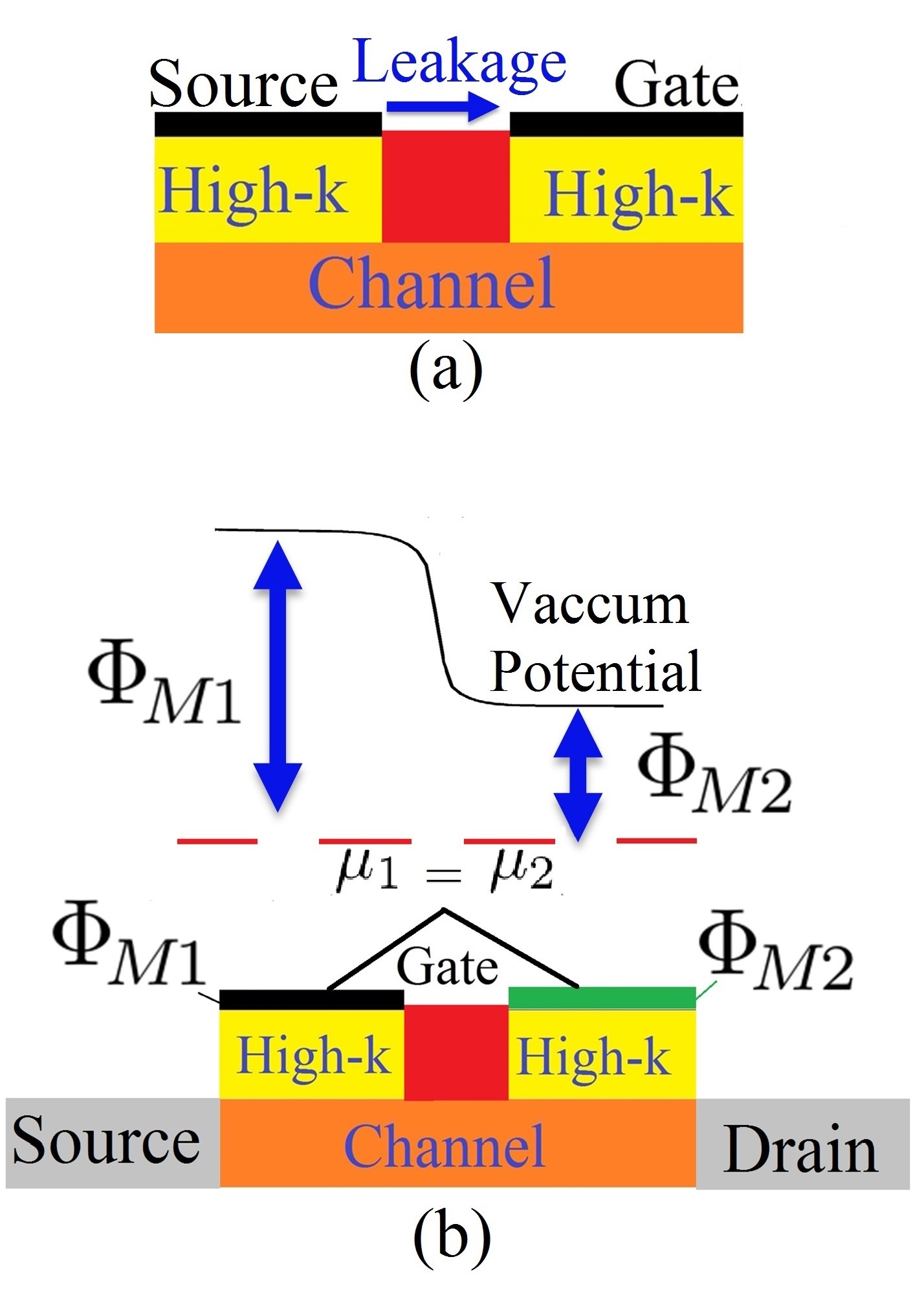}
              % \vspace{-1.5\baselineskip}
              %  \caption{}
                \label{fig:Same_EoT}
        \end{subfigure}%                       
        \begin{subfigure}[b]{0.21\textwidth}
               \includegraphics[width=\textwidth]{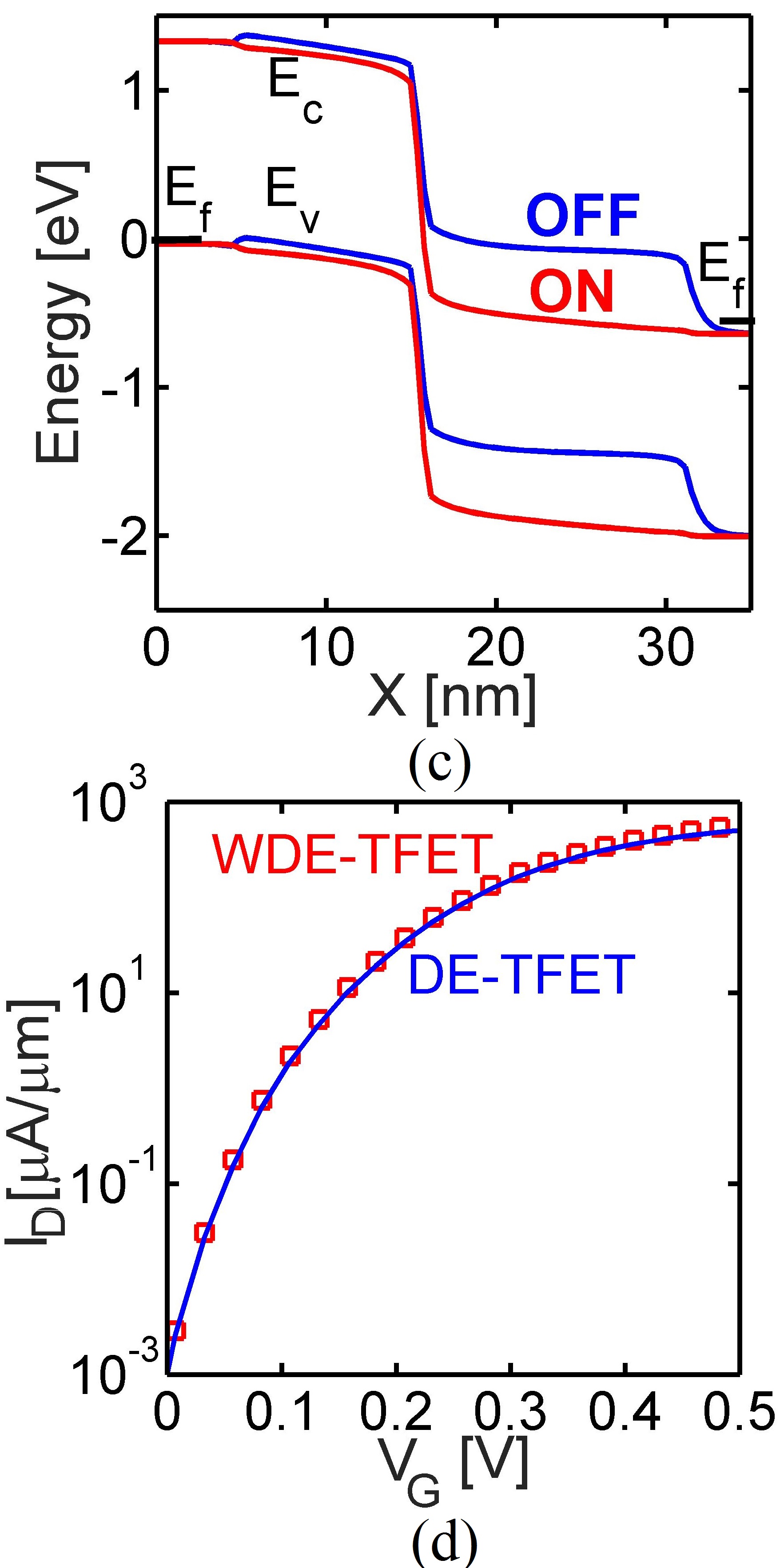}
              % \vspace{-1.5\baselineskip}
              %  \caption{}
                \label{fig:Same_EoT}
        \end{subfigure}%          WDE_TFET                               
        %\vspace{-1.5\baselineskip}
        \caption{ \textbf{a) Leakage problem in DE-TFETs. b) Device structure of workfunction engineered DE-TFET (WDE-TFET). c)Band diagram of WDE-TFET in ON- and OFF-states. d) Tranfer characteristics of WDE-TFET as compared to DE-TFET.} }\label{fig:Fig1}
\end{figure}

From a design perspective, the work function difference between two gate metals should be close to V$_{DD}$+Eg. Such a design leads to a tunneling window of $q V_{DD}$ approximately in the ON-state of the device. Lowering the gate voltage decreases the tunneling window since the potential under metal 2 decreases more than potential under metal 1 (as shown in Fig. 6c) due to larger charge concentration under the contact close to the source region. Fig. 6d shows that WDE-TFET achieves a similar performance as a DE-TFET. 

\section{\textbf{Conclusion}}

The challenge with the sensitivity of steep transistors is revealed using both analytic models and full band quantum transport simulations. Although steep transistors increase the steepness of I-V (response to the gate voltage), they suffer from an increased sensitivity to other design parameters like oxide and channel thicknesses. There is a trade of between performance and sensitivity to design parameters. Although, in III-V channel materials, MOSFETs perform better and are more stable compared to steep transistors, it is shown that for 2D channel materials DE-TFETs present a relatively higher performance and sensitivity as compared with MOSFETs. In comparison with TFETs, DE-TFETs offer lower sensitivity and higher performance. However, the leakage between the contacts limits DE-TFET operationss. To solve this issue, work-function engineering is proposed and the resulting WDE-TEFT is shown to provide a performance similar to DE-TFET.       

%as function of junction thickness
%\clearpage
%\newpage

\bibliographystyle{ieeetr}
\bibliography{thesis}

\end{document}